\documentclass[prb,twocolumn,showpacs,preprintnumbers,superscriptaddress]{revtex4-1}
\usepackage[usenames,dvipsnames]{color}
\usepackage{amssymb} 
\usepackage{times}
\usepackage{graphicx}
\usepackage{graphicx}
\usepackage{dcolumn}
\usepackage{datetime}
\usepackage{soul}
\usepackage{subfigure}
\usepackage[bookmarks, colorlinks=true, plainpages = false,citecolor = red, urlcolor = black, 
filecolor = blue]{hyperref}
\usepackage{multirow} 
\usepackage[english]{babel}
\usepackage{graphicx}
\usepackage{amsmath}
 
\usepackage[english]{babel}
\usepackage{graphicx}
\usepackage{amsmath}
\usepackage[title,titletoc,toc]{appendix}
 
\begin{document}
\title{Gate-tunable valley currents, non-local resistances and valley accumulation 
in bilayer graphene nanostructures}
 
\author{Mohammadhadi Azari} 

\author{George Kirczenow}

\affiliation{Department of Physics, Simon Fraser
University, Burnaby, British Columbia, Canada V5A 1S6}

\date{\today}

\begin{abstract}\noindent
    Using the B\"{u}ttiker-Landauer formulation of transport theory in the linear response regime, the valley currents and non-local resistances of bilayer graphene nanostructures with broken inversion symmetry are calculated. It is shown that  broken inversion symmetry in bilayer graphene nanostructures leads to striking enhancement of the non-local 4-terminal resistance and to valley currents several times stronger than the conventional electric current when the Fermi energy is in the spectral gap close to the energy of Dirac point. The scaling relation between local and non-local resistances is investigated as the gate voltage varies at zero Fermi energy and a power-law is found to be satisfied. The valley velocity field and valley accumulation in four-terminal bilayer graphene nanostructures are evaluated in the presence of inversion symmetry breaking. The valley velocity and non-local resistance are found to scale differently with the applied gate voltage. The unit cell-averaged valley accumulation is found to exhibit a dipolar spatial distribution consistent with the accumulation arising from the valley currents. We define and calculate a {\em valley capacitance} that characterizes the valley accumulation response to voltages applied to the nanostructure's contacts.

\end{abstract}

 
\maketitle
\section{Introduction}
Graphene is a two-dimensional one atom-thick layer of carbon atoms arranged on a honeycomb lattice. Bilayer graphene consists of two electronically coupled single graphene layers. Excellent electrical and thermal conductivity \cite{novoselov2004electric,novoselov2006unconventional,ohta2006controlling,ghosh2010dimensional,balandin2011thermal} at room temperature, high strength, flexibility and transparency\cite{nair2008fine} are properties common to bilayer and monolayer of graphene that make them appropriate candidates for potential applications in future technology. In both monolayer and bilayer graphene the conduction (valence) band has local minima (maxima) called `valleys'. As a consequence, the charge carriers in both materials have a degree of freedom, referred to as the `valley degree of freedom' which is controllable electrically when inversion symmetry is broken. Although the electronic coupling between the two carbon layers that make up the graphene bilayer is relatively weak, it never the less results in important differences between the electronic structures of bilayer and monolayer graphene. Specifically, in contrast to the low-energy linear dispersion of monolayer graphene, the low energy dispersion of bilayer graphene is quadratic. The band structures of monolayer and bilayer graphene both possess interesting topological properties related to the Berry phase and Berry curvature.\cite{XiaoReview} The breaking of inversion symmetry in both materials results in non-zero Berry curvature $\bf{\Omega}_{\bf{k}}$ close to Dirac points. Based on  semiclassical theories of electron transport,\cite{XiaoReview,KL,Chong} a non-zero Berry curvature affects the electron velocity ${\bf v}$ in the presence of an electric field ${\bf E}$ according to 
\begin{equation}\label{vomega}
 {\bf v}_{\bf k}=\frac{1}{\hbar} \frac{\partial \epsilon_{{\bf k}}}{\partial {\bf k}} +\dot{{\bf k}} \times \bf{\Omega}_{\bf{k}}
\end{equation}
where $\epsilon_{{\bf k}}$ is the energy of a Bloch state with wave vector ${\bf k}$ and $\hbar \dot{{\bf k}} = q_e {\bf E}$ in the absence of magnetic fields. Consequently, the Berry curvature $\bf{\Omega}_{\bf{k}}$ can be exploited as a tool to manipulate the valley degree of freedom in these materials when the inversion symmetry is broken since $\bf{\Omega}_{\bf{k}}$ points in opposite directions in the two valleys. In particular, Eq.\ref{vomega} implies a difference in velocity of electrons that belong to different valleys if the inversion symmetry is broken.  This different response of electrons in different valleys to electric fields may be utilized in future valleytronic devices. However, it should be noted that the inversion symmetry breaking (that is crucial for valleytronics based on the Berry curvature mechanism) is achieved in fundamentally different ways for monolayer and bilayer graphene: For monolayer graphene it is imposed by applying a staggered potential that is different at the two atoms of the graphene unit cell. This is done in practice by placing the graphene monolayer on a boron nitride substrate.\cite{gorba} By contrast, for bilayer graphene the inversion symmetry is broken by applying potentials that are uniform throughout each of the monolayers that make up the bilayer but differ between the two monolayers. This is done by dual electrostatic gating.\cite{shima,su}  

Recently, experiments have been carried out, measuring non-local 4-terminal resistances of monolayer graphene on born nitride\cite{gorba} and of dual gate-biased bilayer graphene (BLG) in a Hall bar configuration.\cite{shima,su} In these experiments the inversion symmetries of the materials were broken as described above so as to induce Berry curvature as well as a band gap. Strong enhancement of the non-local (4-terminal) resistance $R_{NL}$ was observed\cite{gorba,shima,su} when the Fermi level passed the energy of Dirac points in the insulating regime. Gorbachev {\em et al.}\cite{gorba} and Shimazaki {\em et al.}\cite{shima} interpreted the enhancement of $R_{NL}$ that they observed in symmetry broken monolayer and bilayer graphene, respectively, as an effect due to valley currents, basing their reasoning on the semiclassical theories of electron transport embodied in Eq.\ref{vomega}. They argued that the anomalous velocity (second term on the right hand side in Eq.\ref{vomega}) generates pure valley currents transverse to the electric current which is flowing within the sample between current contacts. These valley currents then flow between the voltage probes within the sample and result in a potential difference  and enhancement of the non-local resistance.\cite{gorba,shima}

However, the enhanced non-local resistance was observed in the insulating regime,\cite{gorba,shima,su} where the Fermi level was in the energy gap opened in the electronic band structures by the respective symmetry breaking mechanisms.  Therefore the electron transport mechanism involved was quantum tunneling. Hence, as has been discussed for monolayer graphene nanostructures,\cite{george}  the applicability of semiclassical theories of electron transport (Eq.\ref{vomega}) to the observed enhancement of non-local resistance in the insulating regime is questionable because quantum tunneling has no classical analog. In addition, it has been pointed out \cite{george} that, according to B\"{u}ttiker-Landauer  theory\cite{BL}, the Berry curvature mechanism embodied in the anomalous velocity term in Eq.\ref{vomega} cannot affect non-local resistances such as  $R_{NL}$ in the linear response regime. Therefore, any enhancement of $R_{NL}$ in the linear response regime should not be regarded as evidence of valley currents arising from the topological Berry curvature (anomalous velocity) mechanism.  On the other hand, fully quantum mechanical computer simulations based on B\"{u}ttiker-Landauer  theory,  carried out for monolayer graphene nanostructures have shown\cite{george} that in the linear response regime, inversion symmetry breaking gives rise to both strongly enhanced non-local resistances and strong valley currents when the Fermi level is within the energy gap around the Dirac point. The valley currents found in those simulations did not arise from the anomalous velocity term in Eq.\ref{vomega} but were the direct result of electron injection into the monolayer graphene nanostructure.\cite{george}

The previous computer simulations\cite{george} were confined to monolayer graphene nanostructures. 
However, as has been discussed above, the physics of bilayer graphene differs in important ways from that of monolayer graphene with regard to both the electronic structure and the symmetry breaking mechanism. Therefore, whether analogous behavior of the non-local resistance and valley currents should occur in bilayer graphene has remained an open question. From a practical perspective, the gate tunability of the symmetry breaking in bilayer graphene by application of a perpendicular electric field (being not available for monolayer graphene) is an advantage for  bilayer graphene as a potential candidate for applications in future valleytronics. Also while experimental measurements of non-local resistances of symmetry broken bilayer graphene have been reported\cite{shima,su} no fully quantum mechanical calculations  of non-local resistances for this material are available in the literature.  It is therefore of interest to explore  the behavior of valley currents and non-local resistances in bilayer graphene nanostructures theoretically by means of fully quantum mechanical B\"{u}ttiker-Landauer calculations. Such calculations are reported in the present article.

Despite the fundamental differences (described above) between the electronic structures and symmetry breaking mechanisms of bilayer and monolayer graphene, we find some {\em qualitative} similarities between the valley currents and non-local resistances of the two systems. Specifically, we find that in bilayer graphene nanostructures (in common with monolayer nanostructures\cite{george}) inversion symmetry breaking induces strong enhancement of the non-local resistance and strong valley currents transverse to the electric current in the linear response regime when the Fermi level is in the energy gap that is opened by the symmetry breaking. Also, in common with the monolayer case,\cite{george} we find the valley currents in the bilayer nanostructures to be chiral and to be located predominantly near the edges of the bilayer graphene where electrons are injected from electrodes when the Fermi level is in the energy gap. 

However, in the present article we also study quantitatively the scaling of the valley velocity and non-local and local resistances with the dual gate voltage $V_g$ that is responsible for the symmetry breaking in bilayer graphene; such studies are not possible for monolayer graphene where symmetry breaking is imposed instead by the presence of a boron nitride substrate.
We investigate the scaling relations of the local resistance $R_L$ and non-local resistance and valley currents in ballistic nanostructures in the linear response regime at zero temperature as the gate voltage varies for $E_F=0$. We find the power law $R_{NL}\propto R_L^\alpha$  to be satisfied with $\alpha=2.19$ in our ballistic bilayer nanostructures, whereas $\alpha \sim 2.77$ has been observed experimentally in bilayer graphene in the diffusive regime.\cite{su} While theoretical predictions for $\alpha$ have not been available previously for bilayer graphene,  $\alpha=3$ has been predicted for diffusive spin Hall systems.\cite{Abanin} 
Interestingly, we find the normalized valley current to scale linearly with gate voltage for low gate voltages while the non-local resistance scales quadratically starting from a non-zero value at zero gate voltage.

We also introduce and calculate the {\em valley accumulation} and {\em valley capacitance} associated with valley currents. While there have been previous theoretical studies of valley currents induced in graphene in various ways,\cite{george,XiaoPRL,Rycerz,Nakanishi,Gunlycke,Oka,Abergel,Golub,Costa} whether and to what
extent the valley currents result in valley accumulation of electrons, i.e., in differing electron populations of the different valleys, has remained unclear. To our knowledge, there has been no estimate of the valley accumulation or of its spatial distribution resulting from valley currents in any graphene device. This is despite the central role that valley accumulation is expected to play in future valleytronic devices whose operation, by definition, depends on imbalances between the electron populations of the different valleys. We therefore investigate the valley accumulation of electrons associated with valley currents in bilayer graphene nanostructures and report our results here. We find the valley accumulation to typically have opposite signs on the two carbon atoms of the unit cell of each graphene monolayer of the bilayer. After averaging over the unit cell, we find the cell-averaged valley accumulation to exhibit a dipolar spatial distribution. The dipole is oriented along the overall direction of the valley current flow, consistent with the valley current giving rise to the dipolar accumulation. In order to develop an intuitively appealing figure of merit for the magnitude of the response of the valley accumulation to voltages applied to the contacts of
graphene nanostructures in the linear response regime we define a {\em valley capacitance} and evaluate it for a bilayer graphene nanostructure with broken inversion symmetry.  

The present model of bilayer graphene with broken inversion symmetry, the formalism of B\"{u}ttiker-Landauer theory, the Lippmann-Schwinger equation and their application in calculations of the non-local resistance, valley currents and valley accumulation are described in Sec.\ref{MF}. Our results are presented in Sec.\ref{Results}. We summarize and discuss our main conclusions in Sec.\ref{Discussion}.  Technical details of the method of solution of the Lippmann-Schwinger equation that yields our calculated scattering amplitudes that enter the B\"{u}ttiker equations that yield our calculated non-local resistances, and also the wave functions of our transport states are summarized in Appendix A. The calculation of the valley-projected electronic states is described in Appendix B. 

\section{Model and Formalism} 
\label{MF}
\begin{figure}[t!]
\centering
\includegraphics[width=1.0\linewidth]{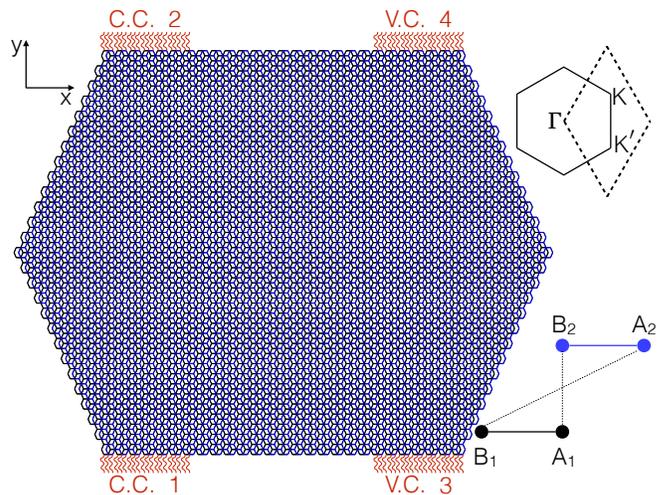}
\caption{(Color online) 4-terminal BLG nanostructure with armchair edges. The bottom (top) layer is shown in black (blue). Each contact is composed of 40, semi-infinite 1D ideal leads (shown by red wavy lines) that are attached to both layers and connect the nanostructure to the reservoirs. The electric current  flows through current contacts 1 and 2 while there is no net electric current entering or leaving the voltage contacts 3 and 4. In non-local resistance studies the potential difference is measured between contacts 3 and 4. Upper right inset: Two types of first Brillouin zone of bilayer graphene, hexagonal (solid) and rhombic (dotted). Lower right inset: Side view of 4 carbon atoms of a unit cell in bilayer graphene in AB stacking. The inversion symmetry point in the unit cell is the intersection of dotted lines.   
}
\label{nanostructure} 
\end{figure}

To describe the bilayer graphene nanostructure with AB stacking of the two honeycomb lattice layers, the nearest neighbour tight-binding Hamiltonian $H_{BLG}$ is employed such that
\begin{multline}
H_{BLG}=\sum_n\epsilon_na_n^{\dag}a_n-\sum_{\langle n,m\rangle}t_{nm}(a_n^{\dag}a_m+H.c.)+\\
\sum_{\langle n_1,m_2\rangle}t_{n_1m_2}(a_{n_1}^{\dag}a_{m_2}+H.c.)
\label{Hamiltonian}
\end{multline}
where $\epsilon_n$ is the on-site energy, $t_{nm}$ = t = 2.7 eV is the nearest neighbour hopping amplitude between the $p_z$ orbitals of carbon atoms belonging to the same graphene monolayer. $t_{n_1m_2}$ = 0.1t is the hopping amplitude between the $p_z$ orbitals of nearest neighbour  carbon atoms belonging to different graphene monolayers, such as carbon atoms $A_1$ and $B_2$ in the lower right inset of Fig.\ref{nanostructure}. It should be noted that, the interlayer coupling (the last term on the right hand side of Eq.\ref{Hamiltonian}) is absent in the tight-binding Hamiltonian of monolayer graphene and is responsible for the difference between the electronic band structures of monolayer and bilayer graphene. To break the inversion symmetry of the structure, we have chosen $\epsilon=+V_g/2$ on the atoms of the top graphene layer and $\epsilon=-V_g/2$ on the atoms of the bottom layer to model the effects of the perpendicular electric field in the Sui {\em et al.}\cite{su}  and Shimazaki {\em et al.}\cite{shima} experiments. In contrast to monolayer graphene samples that require a hBN substrate to break the inversion symmetry and introduce a band gap around the Dirac point\cite{gorba}, the application of a perpendicular electric field in bilayer graphene nanostructures enables the investigation of gate tunability of non-local topological transport. Since, precise alignment of the crystal axes of the monolayer graphene sample and hBN substrate is required during the fabrication of monolayer graphene valleytronic devices, bilayer graphene nanostructures (that do not require this alignment) may be more attractive as candidates for applications in the future technology. 

A variety of different methods for calculating quantum transport coefficients within tight-binding formalisms
have been developed and applied in the literature. They include Landauer mode counting applied to calculated quasi-one-dimensional 
band structures,\cite{Geo,Peres} recursive Green's function techniques,\cite{LF,MacK,Ando,Xu} non-equilibrium Green's function methods,\cite{Datta,Lake} stabilized transfer matrix methods,\cite{Usuki,Akis,KJKGSFD} and solution of the Lippmann-Schwinger equation,\cite{Nonoyama,GRings,Dalgleish1} among others. In the present work we chose the
Lippmann-Schwinger approach for the following reasons: It is very flexible, lending itself well to calculations of transport in the linear response regime for tight-binding models of nanostructures of many different materials, with arbitrary geometries. This versatility has made possible its application to theoretical studies of electron transport  in 2D semiconductor nanostructures,\cite{Nonoyama} in disordered metal nanostructures \cite{GRings}, in ferromagnetic atomic contacts, \cite{Dalgleish1}in monolayer graphene nanostructures,\cite{george} in molecules bridging transition metal electrodes\cite{Dalgleish2} and gold electrodes,\cite{AFG} in molecular spin current rectifiers,\cite{Dalgleish4} in arrays of molecules on silicon,\cite{Piva3} in electrochemically gated protein nanowires,\cite{Cardamone1} in single molecule nanomagnets bridging metal electrodes,\cite{Renani1} in Fe/GaAs spin valves,\cite{Majumder} in scanning tunneling microscopy of molecules,\cite{Buker} in molecular electroluminescence,\cite{Buker1}in ballistic electron spectroscopy of buried molecules,\cite{George2}  in vibrational spectroscopy of molecular junctions,\cite{Firuz1}  and others. The Lippmann-Schwinger equation is readily applicable to calculations of transport in nanostructures with arbitrary numbers of electrical contacts\cite{george} when combined with the B\"{u}ttiker-Landauer formalism\cite{BL}. It yields exact numerical solutions of the transport coefficients for the tight binding model described by Eq.\ref{Hamiltonian} in the linear response regime. Thus it is well suited to the present study of non-local resistances and valley currents in four-terminal bilayer graphene nanostructures.

According to B\"{u}ttiker-Landauer theory\cite{BL} at zero temperature in the linear response regime, the current $I_i$ in each contact $i$ in a multiterminal nanostructure is related to the electrochemical potential of that contact $\mu_i$ and other contacts $\mu_j$ by 
\begin{equation}\label{buttiker}
I_i=\frac{q_e}{h}(N_i\mu_i-R_{ii}\mu_i-\sum_{j\neq i}T_{ij}\mu_j),
\end{equation}
where $N_i$ is the total number of modes supported by contact $i$, $T_{ij}$ is the electron transmission probability from contact $j$ to contact $i$, and $R_{ii}$ is the electron reflection probability from the nanostructure for contact $i$. In this study, each contact is represented by a group of semi-infinite one-dimensional tight-binding leads with one orbital per site, as in many previous theoretical studies of quantum transport.\cite{Buker,Buker1,Firuz1,Cardamone1,Piva3,Dalgleish1,Dalgleish2,Dalgleish4,George2,Renani1,AFG,Majumder} These ideal leads (represented by wavy lines in Fig.\ref{nanostructure}) are attached to edge sites of both graphene layers. 

In order to calculate the $T_{ij}$ coefficients of B\"{u}ttiker-Landauer theory,  we have solved the Lippmann-Schwinger equation 
\begin{equation}\label{lippmann}
|\psi^l\rangle=|\phi_{\circ}^l\rangle+G_{\circ}(E)W|\psi^l\rangle,
\end{equation}
as is described in Appendix A. Here $|\phi_{\circ}^l\rangle$ is an eigenstate of the $l^{th}$ lead that is decoupled from the BLG nanostructure, $G_{\circ}(E)$ is the sum of the Green's functions of the BLG nanstructure and leads when they are decoupled, and $|\psi^l\rangle$ is the scattering eigenstate of the coupled system. $W$ is the coupling between the BLG nanostructure and the ideal leads, i.e.,
\begin{equation}
W=-\sum_nt(b_n^{\dag}a_n+H.c.)
\end{equation}
where $b_n^{\dag}$ is the electron creation operator at a lead site attached to the nanostructure, $a_n$ is the electron annihilation operator at the BLG site attached to the corresponding lead, and the hopping amplitude $t$ is assumed to be the same as the hopping between the $p_z$ orbitals of in-plane nearest-neighbour atoms of the BLG nanostructure. Having evaluated scattering states $|\psi^l\rangle$, the coefficients $T_{i,j}$ that enter B\"{u}ttiker-Landauer theory (Eq.\ref{buttiker}) are given by 
\begin{equation}
T_{ij}(E)=\sum_{l,p}|t_{lp}^{ij}|^2 \frac{v_l^i}{v_p^j},
\end{equation}
where $t_{lk}^{ij}$ is the quantum transmission amplitude of an electron transmitted from the $k^{th}$ lead of contact $j$ to the $l^{th}$ lead of contact $i$ at energy E obtained from the scattering states $|\psi^l\rangle$. $v^{i(j)}_{l(p)} = \frac{1}{\hbar} \frac{\partial\epsilon}{\partial k}$ is the electron velocity in the 1-D semi-infinite lead $l(p)$ of contact $i(j)$ at energy E, and $\epsilon$ are the energy eigenvalues of the tight-binding Hamiltonian of the semi-infinite ideal lead.
To calculate the 4-terminal non-local resistance $R_{NL}$ of a BLG nanostructure, following the calculation of electron transmission probabilities $T_{ij}$ at the Fermi energy $E_\text{F}$, the B\"{u}ttiker equations Eq.\ref{buttiker} are solved. Then
\begin{equation}\label{nonlocalres}
R_{NL}=\frac{\Delta V_{3,4}}{I_{1,2}},
\end{equation} 
where $I_{1,2}$ is the electric current flowing between the current contacts 1 and 2 and $\Delta V_{3,4}$ is the potential difference between the voltage contacts 3 and 4. Here the potential difference $\Delta V$ and electrochemical potentials $\mu$ appearing in the B\"{u}ttiker equations Eq.\ref{buttiker} are related by $\Delta V=\Delta \mu/q_e$.

In order to estimate the valley currents induced  in a BLG nanostructure in response to the electrochemical potential differences between the contacts, the scattering states $|\psi^l\rangle$ of electrons incident from each lead $l$ at energy E are evaluated by solving the Lippmann-Schwinger equation Eq.\ref{lippmann}. Then the electron scattering states $|\psi^l\rangle$ are projected onto the Bloch states of BLG. As in Ref.\onlinecite{george}, the Bloch state is assumed to belong to valley $K (K')$ if its wave vector lies within the the upper (lower) half of the rhombic Brillouin zone represented by dotted lines in the upper right inset of Fig.\ref{nanostructure}. Futher details of the projection method are described in Appendix B. 

The $\eta$-component of the velocity operator for electrons in BLG nanostructure is 
\begin{equation}\label{velocity}
v_{\eta}=\frac{1}{i\hbar}[\eta,H_{BLG}],
\end{equation} 
where $\eta=\sum_n\eta_n a_n^{\dag}a_n$ and $\eta_n$ is the $\eta$ coordinate of atomic site $n$. Then, the valley velocity is expressed in terms of the expectation values of the velocity operator Eq.\ref{velocity} with respect to projected states $|\psi_K^l\rangle$ and $|\psi_{K'}^l\rangle$ of valley $K$ and $K'$, respectively. For bilayer graphene nanostructures with multiple contacts $i$, each at its own electrochemical potential $\mu_i$, we define the valley velocity as an average over the contributions arising from  the different contacts weighted according to the electrochemical potential differences between the contacts.\cite{george} In particular, the weighted average velocities of electrons in valley $K$ and $K'$ are defined by 
\begin{equation}
v_{K(K')\eta}=\frac{\sum_{l,i}\frac{1}{N_{li}}\langle\psi^{li}_{K(K')}|v_{\eta}|\psi^{li}_{K(K')}\rangle\Delta\mu_i}{\sum_{l,i}\Delta\mu_i},
\label{vval}
\end{equation} 
where $N_{li}$ is an appropriate wave function normalization factor and $\Delta\mu_i$ is the electrochemical potential difference between contact $i$ and the lowest electrochemical potential of all of  the contacts. Having evaluated $v_{K\eta}$ and $v_{K'\eta}$, the weighted valley velocity is defined here as 
\begin{equation}
v_{\eta}^{val}=v_{K\eta}-v_{K'\eta}.
\label{valvel}
\end{equation}

As in Ref.\onlinecite{george} for monolayer graphene, the valley velocity field for the BLG nanostructure is defined by expressing the expectation value of the valley velocity given by Eq.\ref{valvel} and \ref{vval} as a sum of contributions of nearest neighbour pairs of C atoms in each layer and assigning each such contribution to the midpoint of the respective atomic pair. 

From the perspective of potential valleytronic applications it is also important to know the valley accumulation and its spatial distribution in the device. For an electron in an appropriately normalized state $|\psi^l\rangle$, the valley accumulation in that state at atomic site $n$ with carbon atom $p_z$ orbital $|n\rangle$ is defined by $a^l_n = |\langle n|\psi^l_K\rangle|^2-|\langle n|\psi^l_{K'}\rangle)|^2$. For a nanostructure with multiple contacts in the linear response regime, the total valley accumulation $A_n$ induced at atomic site $n$ by changes $\Delta\mu_i$ in the electrochemical potentials $\mu_i$ of contacts $i$ is then given by
\begin{equation}\label{accumulation}
A_n=\frac{1}{2\pi}\sum_{l,i}(|\langle n|\psi^l_K\rangle|^2-|\langle n|\psi^l_{K'}\rangle)|^2)\frac{\partial \zeta^l}{\partial E}\Delta\mu_i,
\end{equation}
where it is assumed that the scattering state $\psi^l$ originates in a semi-infinite 1D ideal tight-binding chain where it is normalized so that on site $m$ of the chain $\langle m|\psi^l\rangle = e^{i \zeta^l m} + r^l e^{-i \zeta^l m}$ where $r^l$ is the reflection amplitude of the incoming state $|\psi^l\rangle$ from the nanostructure back into ideal lead $l$, and $E$ is the energy eigenvalue corresponding to state $\psi^l$.

It is then natural to define valley accumulation susceptibilities per carbon atom as
\begin{equation}\label{accumulationci}
C_{ni} = e^2 \frac{\partial A_n}{\partial \Delta\mu_i}
\end{equation}
where the factor $e^2$ is introduced so that $C_{ni}$ has the units of capacitance.
Henceforth we shall refer to the $C_{ni}$ as ``partial valley capacitances." For a 
multiterminal device in the linear response regime we characterize 
the total valley accumulation due to the contributions of all of the contacts 
by a total
valley capacitance defined by
\begin{equation}\label{Ctotal}
C_{n} = e^2 \frac{A_n}{\Delta\mu_\text{max}}
\end{equation}
where $\Delta\mu_\text{max}$ is the largest of the $\Delta\mu_i$ over all of the
contacts $i$.

We find the valley capacitances to often have differing signs at the different
atoms in the unit cell, and therefore to better characterize the overall magnitude
of the valley accumulation we define an averaged valley capacitance by
\begin{equation}\label{averagecap}
C_\text{av}(x,y)=\frac{C_l+C_k}{2}
\end{equation}     
where $(x,y)=[(x_l+x_k)/2,(y_l+y_k)/2]$, and $C_l$ and $C_k$ are the total valley 
capacitances calculated for two carbon atoms belonging to the same graphene 
layer and in the same unit cell. Inspection of spatial maps of these cell-averaged 
capacitances facilitates developing an understanding of the relationship between
valley currents and valley accumulation, as will be seen below.  
 \begin{figure}[t!]
\centering

\includegraphics[width=1.0\linewidth]{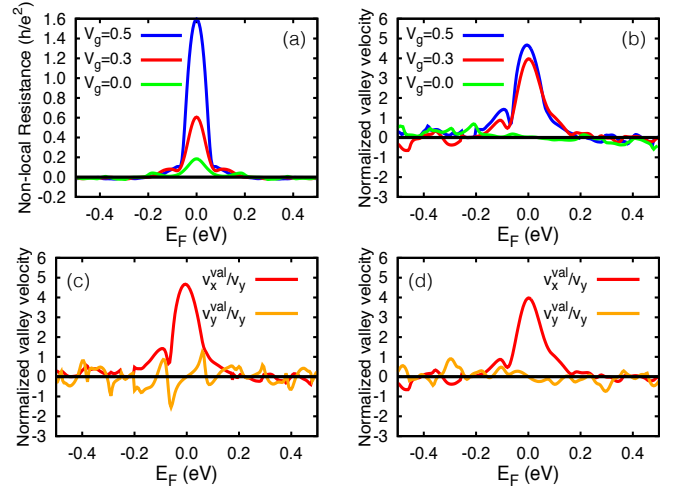}
\caption{(Color online)(a) Calculated non-local resistance $R_{NL}$ (Eq.\ref{nonlocalres}) of the nanostructure of Fig.\ref{nanostructure} in the linear response regime at zero temperature for different values of the gate voltage $V_g=0$ eV (green), $V_g=0.3$ eV (red), and $V_g=0.5$ eV (blue) as a function of Fermi energy $E_F$. (b) Normalized valley velocity $v_x^{val}/v_y$ of the BLG nanostructure for different values of the gate voltage as a function of Fermi energy when the net current flows between current contacts 1 and 2. (c) Comparison of the $x$ and $y$ components of the normalized valley velocity as a function of Fermi energy at $V_g=0.5$ eV.  (d) Comparison of the  $x$ and $y$ components of normalized valley velocity as a function of Fermi energy at $V_g=0.3$ eV.
}
\label{RNLandValcur} 
\end{figure}

\section{Results} 
\label{Results}
The results of our calculation of the four-terminal non-local resistance obtained by the application of B\"{u}ttiker-Landauer theory (Eq.\ref{buttiker}) at zero temperature in the linear response regime are shown in Fig.\ref{RNLandValcur}(a) as a function of the Fermi energy for different values of the gate voltage. As is seen in Fig.\ref{RNLandValcur}(a), enhancement of the non-local resistance $R_{NL}$ occurs near the energy of Dirac point $E_F=0$. The breaking of inversion symmetry of the BLG nanostructure by application of a potential $+V_g/2$ to the top layer and $-V_g/2$ to the bottom layer leads to the striking enhancement of the non-local resistance. As the gate voltage increases the maximum value of $R_{NL}$ which occurs at $E_F=0$ also increases implying that non-local  electron transport in the BLG nanostructure is gate-tunable. Furthermore, as can be seen in Fig.\ref{RNLandValcur}(a), when the Fermi energy is well away from the energy of Dirac point, the symmetry breaking does not result in large non-local resistances $R_{NL}$. 

    The results for the normalized valley velocity $v_\eta^{val}/v_y$  as a function of the Fermi energy for different values of gate voltage $V_g$ are presented in Fig.\ref{RNLandValcur}(b)-(d). The computed weighted valley velocities in the $x$ and $y$-directions are normalized by the weighted electron velocity in the $y$-direction $v_y$. [Since the net electric current flows between the current contacts 1 and 2 (shown in Fig.\ref{nanostructure}) the weighted average velocity of electrons in the x-direction $v_x$ is near zero within numerical error, as expected since the net electric current points in the y-direction.]  In Fig.\ref{RNLandValcur}(b) the computed normalized $x$-component of valley velocity or valley current ($v_x^{val}/v_y=I_x^{val}/I$) is shown for different values of the gate voltage where $I$ is the total electric current flowing through the nanostructure. It is seen that the $x$-component of valley current $I_x^{val}$ peaks when the Fermi energy passes the energy of Dirac point. It should be pointed out that when the inversion symmetry breaking is absent ($V_g=0$) the evaluated valley current in the $x$-direction is zero at $E_F=0$, while its value exceeds the value of electric current flowing through the nanostructure in the presence of inversion symmetry breaking for all of the values of $V_g\neq0$ that are shown. A comparison of the $x$ and $y$-components of normalized valley velocities is presented in Fig.\ref{RNLandValcur}(c) and (d). Based on these results the $y$-component of valley current approaches zero as the Fermi energy passes the energy of Dirac point. To summarize, the valley currents flow through the BLG nanostructure mainly in the $x$ direction. Also, the $x$ and $y$-components of valley current are relatively small in the presence or absence of inversion symmetry breaking if the Fermi energy is well away from of Dirac point.
 \begin{figure}[t!]
\centering
\includegraphics[width=1.0\linewidth]{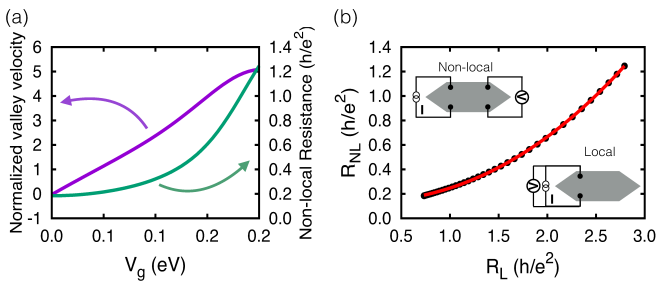}
\caption{(Color online) (a) The normalized valley velocity ($v_x^{val}/v_y$) (purple line) and non-local resistance $R_{NL}$ (green line) of the bilayer graphene nanostructure as a function of gate voltage at zero Fermi energy $E_F=0$. (b) The scaling relation between the local and non-local resistance as the gate voltage varies from 0 eV to 0.2 eV. Red line is the power law $R_{NL}\sim R_L^\alpha$ fitted to the simulation data. Upper left inset: The configuration of non-local resistance measurements. Lower right inset: The configuration of local resistance measurements.}
\label{scaling} 
\end{figure}

    For comparison, we have also calculated the non-local resistance and valley current in the linear response regime for a model nanostructure whose dimensions (including the widths of the contacts) are one half of those of the nanostructure considered in Figs. 1 and 2. The simulation results, show that when the gate voltage $V_g=0.3 eV$,  the non-local resistance of the smaller nanostructure is smaller by a factor of $\sim 16$ at $E_F=0$. However, the calculated normalized valley current in the $x$-direction is smaller by a factor of $\sim 2.5$ for the smaller sample. The valley current in the $y$-direction is still equal to zero when the Fermi level passes the energy of the Dirac point.  

    Returning to the bilayer graphene nanostructure in Fig.\ref{nanostructure}, in Fig.\ref{scaling}(a) we show the gate-tunability of the valley velocity in the $x$-direction ($v_x^{val}/v_y$) and of the non-local resistance at zero Fermi energy (the energy at which the maximum values of the valley current and non-local resistance occur) as the gate voltage $V_g$ varies. As can be seen, the valley current (purple line) increases from zero (when the inversion symmetry is absent, i.e., $V_g=0$) to a value of five times larger than the electric current. The normalized valley current increases linearly as the gate voltage increases for low values of the gate voltage (i.e. $0<V_g<0.1 eV$). By contrast, the computed non-local resistance $R_{NL}$ (green line) of the BLG nanostructure increases quadratically (from a non-zero value at $V_g=0$) as a function of gate voltage. Deviations from these scaling relations of valley current and non-local resistance occur for higher values of the gate voltage (i.e. $V_g>0.1eV$). The relationship between the local and non-local resistances is investigated in Fig.\ref{scaling}(b) as the gate voltage varies from 0 eV to 0.2 eV at zero Fermi energy. The local resistance $R_{L}$ is computed when the electric current flows through the nanostructure between current contacts 1 and 2 and the potential difference is measured between the same contacts ($R_L=\Delta V_{1,2}/I_{1,2}$). According to Fig.\ref{scaling}(b), the power-law relation $R_{NL}\sim R_L^\alpha$  is satisfied with $\alpha=2.19$. Deviation of $\alpha$ from the experimental results $\alpha=3$\cite{shima} and $\alpha=2.77$ \cite{su} may result from the difference between ballistic transport system investigated in this paper and diffusive transport in the experiments. As reported\cite{su}, the power $\alpha$ differs in different samples as a consequence of disorder in the samples.  
    
 \begin{figure*}[t]
\centering
\includegraphics[width=1.0\linewidth]{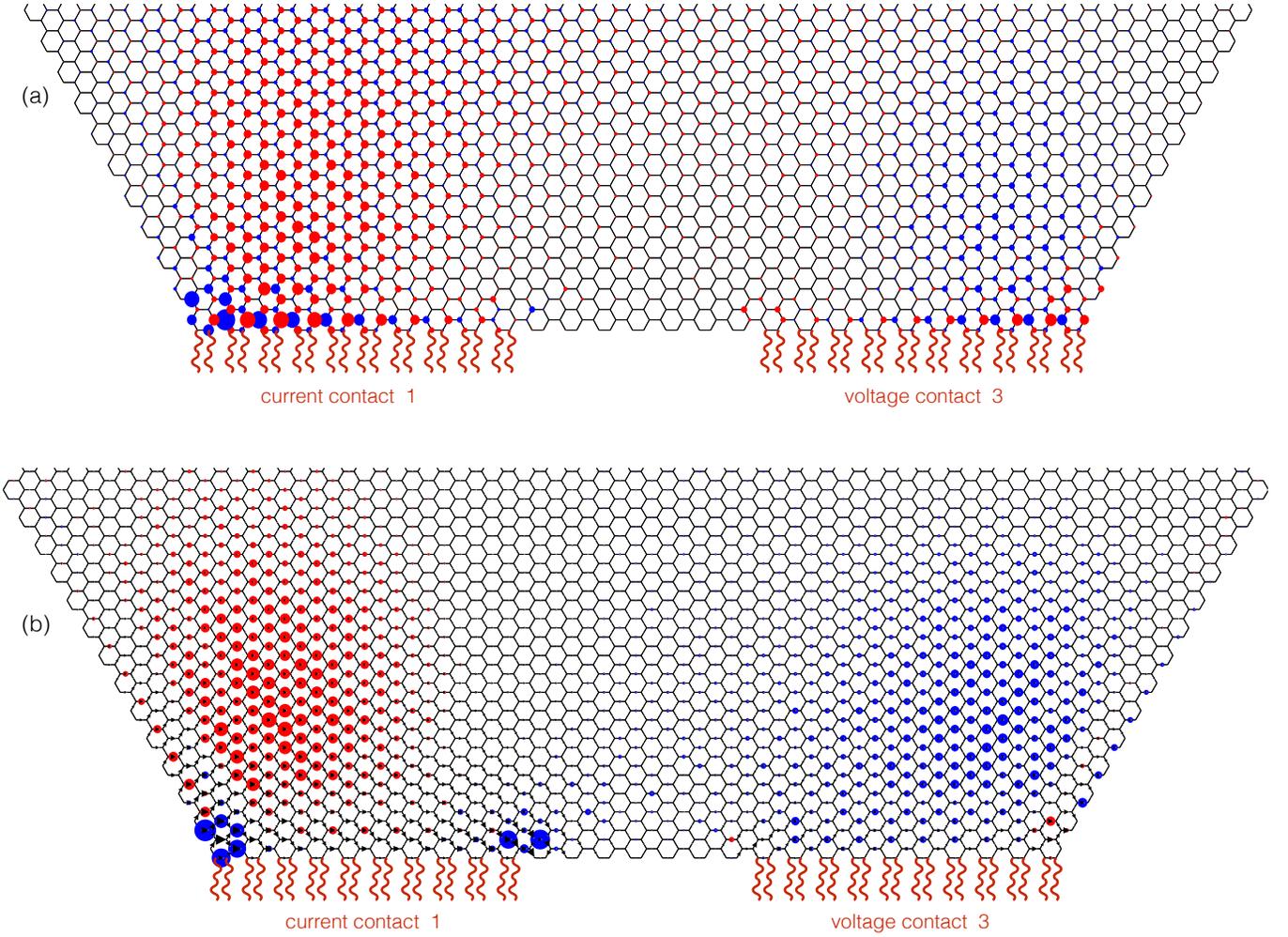}
\caption{(Color online) The calculated total valley capacitance, valley velocity field, and unit cell-averaged valley capacitance of the bottom layer of a bilayer graphene nanostructure when the net electric current flows between current contact 1 and 2 at $V_g=0.15$ eV and $E_F=0$. (a) The total valley capacitance (Eq.\ref{Ctotal}) is represented by blue (red) discs when the on-site valley accumulation is positive (negative). (b) The valley velocity field (black arrows) and average valley capacitance (red and blue discs), Eq.\ref{averagecap}.}
\label{valvelfield}    
\end{figure*} 

The total valley capacitance (Eq.\ref{Ctotal}) map of the bottom layer of the BLG nanostructure is presented in Fig.\ref{valvelfield}(a) for the net current flowing between current contacts 1 and 2. The blue discs indicate a positive total valley capacitance $C_n$ and red discs represent negative $C_n$. As can be seen in Fig.\ref{valvelfield}(a), the valley capacitance is stronger on sites close to the contacts from which electrons enter the nanostructure even if no net current flows through that contact. It should also be noted that the total valley capacitance in the top layer (not shown) exhibits similar behaviour to that for the bottom layer of the BLG nanostructure. 

In Fig.\ref{valvelfield}(b) the spatial distribution of the valley velocity and average valley capacitance (Eq.\ref{averagecap}) in each unit cell of the bottom layer for broken inversion symmetry ($V_g=0.3 eV$) is represented by arrows (black) and discs (red and blue), respectively. As can be seen the valley current is chiral and directed from left to right along the lower boundary of the nanostructure, while along the opposite boundary (not shown) the valley current is also chiral but weaker and travels in opposite direction, from right to left. The overall directions of the valley velocities along the lower and upper boundaries of the BLG nanostructure reverse if the sign of $V_g$ is changed. Apart from at a few exceptional sites near the ends of contacts 1 and 3, the cell-averaged valley capacitance in Fig.\ref{valvelfield}(b) exhibits a dipolar distribution, being negative (red) on the left and positive (blue) on the right.  The maximum absolute value of the cell-averaged valley capacitance $|C_{av}(x,y)|$ is $\sim0.12 \times10^{-22}$ Farad/atom. For comparison, a classical parallel plate capacitor with a 3.1 \AA~plate spacing would have a conventional capacitance of $\sim7 \times10^{-22}$ Farad per  atomic area of graphene.

The dipolar character of the average valley capacitance in Fig.\ref{valvelfield}(b) [the negative (red) region on the left and  positive (blue) region on the right] has physical (and potentially practical) significance. It indicates that the valley current that flows from left to right in fact transports the valley degree of freedom across the nanostructure and results in a significant valley imbalance between these two regions of the nanostructure, despite the fact that the electron crystal momentum is not conserved when an electron scatters from a boundary of the nanostructure. 

The computed average valley capacitance is negligible in the absence of inversion symmetry breaking and grows as the applied gate voltage increases. It should be noted that the average valley capacitance in the unit cell and valley velocity field in the top layer are similar to those in the bottom layer that are shown in Fig.\ref{valvelfield} (b). 

 \section{Conclusions} 
 The present study has investigated the non-local resistance, valley currents and average valley capacitance of four-terminal bilayer graphene nanostructures from the perspective of scattering theory in the insulating regime. The calculations have been carried out in the limit where the effect of the electric field that drives electrons has been sent to zero. As a consequence, in the linear response regime, the interpretation of the valley currents by application of the anomalous velocity term is not appropriate. The valley currents calculated here within the formalism of scattering theory are not due to the acceleration of electrons in an electric field when the inversion symmetry is broken (non-zero Berry curvature), but result from non-adiabatic injection of electrons into the BLG nanostructure through the contacts. The valley currents are maximal in the insulating regime where the Fermi energy is in a spectral gap close to the energy of Dirac point, so that electron transport is only possible by quantum tunneling. Consequently, the calculated valley currents are strongest in the BLG nanostructure with broken inversion symmetry close to the contacts from which electrons enter the nanostructure even if there is no net electric current passing through those contacts (the case of voltage contacts of the BLG nanostructure). For the Fermi energies near  the energy of the Dirac point, the valley currents are several times larger than the electric current passing through the current contacts when the inversion symmetry is broken. On the other hand, if the Fermi energy is well away from the Dirac point, the impact of inversion symmetry breaking on the overall magnitude of the valley currents is not significant. Furthermore, mapping the average valley capacitance calculated in the BLG nanostructure with broken inversion symmetry reveals a dipolar distribution that results from the presence valley currents so that the average valley capacitance increases in the direction of valley currents. The evaluated scaling relation between local and non-local resistances as the gate voltage varies, shows that the diffusive power-law is satisfied in the ballistic linear response regime although with a modified exponent. The valley currents and average valley accumulation in bilayer graphene are electrically controllable by means of a perpendicular electric field which is an advantage for the application of bilayer graphene over monolayer graphene nanostructures. It should be noted that the valley currents and average valley capacitances in both layers of the bilayer graphene have similar characteristics.
 
 This research was supported by NSERC, Westgrid, CIFAR and Compute Canada.        
\label{Discussion}

\numberwithin{equation}{section}
\begin{appendices}
\section{Solution of the Lippmann-Schwinger equation}
In this Appendix we discuss the evaluation of the decoupled Green's function $G_{\circ}(E)$ that enters the Lippmann-Schwinger equation Eq.\ref{lippmann}, the method of solution of the Lippmann-Schwinger equation in the tight-binding formalism and how this solution relates to the B\"{u}ttiker equations and the electronic transport states.
 
The Green's function of each decoupled contact represented by a group of semi-infinite one-dimensional leads depicted in Fig.\ref{nanostructure}, is the sum of the Green's functions of the individual decoupled leads associated with that contact. The energy of an electron in a one-dimensional lead with one atomic orbital per site in the tight-binding model is $E_k=\epsilon+2\tau cos(ka)$, where $\epsilon$ is the onsite energy (equal to the onsite energy of the carbon atom attached to that lead), $\tau$ is the hopping amplitude between the nearest neighbors ($|\tau| =t$ in the bilayer graphene tight-binding Hamiltonian Eq.\ref{Hamiltonian}), and $a$ is the nearest neighbor spacing between the atomic sites in one-dimensional lead. Therefore, the electron wave function in the $l^{th}$ lead when it is decoupled from the graphene nanostructure has the form
\begin{equation}
|\phi_{\circ}\rangle = \frac{1}{\sqrt{2N}}\sum_{n_l=1}^N(e^{ikn_la}-e^{-ikn_la})|n_l\rangle,
\end{equation}
where, N is the total number of atoms (assumed to be very large)  in the one-dimensional lead, and $|n_l\rangle$ is the $n^\text{th}$ atomic orbital state in that lead. Then, the decoupled Green's function of the $l^{th}$ lead is:
\begin{equation}
\begin{aligned}
G_{\circ}^l(E)&=  \frac{1}{E-H_{\circ}^l+i\delta} \\
                     &=\sum_{n_l {n_l}'}(G_{\circ})^l_{n_l {n_l}'}|n_l\rangle\langle {n_l}'| 
\end{aligned}
\end{equation}
where $H_{\circ}^l$ is the Hamiltonian of the $l^\text{th}$ decoupled 1D lead and $(G_{\circ})^l_{n_l {n_l}'}=\langle n_l |G_{\circ}^l(E)|{n_l}'\rangle$. For our purposes this decoupled Green's function needs
to be evaluated only at an end site $(n_l=\pm1)$ where the $l^{th}$ lead will ultimately be attached to the BLG nanostructure. Its
corresponding matrix element can be written as: 
\begin{equation}
(G_{\circ})^l_{\pm1,\pm1}=\frac{1}{2\pi}\int_0^{2\pi}dk\frac{1-e^{2ik}}{(E-(\epsilon+2\tau cos(k))+i\delta)}.
\label{Green}
\end{equation}
By application of residue theorem, the Green's function of the decoupled semi-infinite one-dimensional $l^{th}$ lead is 
\begin{equation}
(G_{\circ})^l_{\pm1,\pm1}=\frac{i}{2\tau}\frac{(1-e^{2ika})}{sin(ka)}.
\end{equation} 
It should be noted that in derivation of Eq.\ref{Green} we have used the conversion of summation to the integral as $\sum_k\longrightarrow\frac{L}{2\pi}\int_{-\frac{\pi}{a}}^{\frac{\pi}{a}}dk$. Then, the Green's function of decoupled bilayer graphene nanostructure can be written as 
\begin{equation}
G_{\circ}^{BLG}(E)=\sum_\alpha\frac{|\alpha\rangle\langle \alpha|}{E-\epsilon_\alpha+i\delta}
\label{green2}
\end{equation}
where $\epsilon_\alpha$ are the energy eigenvalues and $|\alpha\rangle$ are the corresponding eigenstates of the BLG Hamiltonian Eq.\ref{Hamiltonian} decoupled from the leads. $\epsilon_\alpha$ and $|\alpha\rangle$ are calculated numerically by diagonalizing
the BLG Hamiltonian.

To evaluate the quantum transmission amplitude of an electron $t_{lk}^{ij}$, transmitted from the $k^{th}$ lead of contact $j$ to the $l^{th}$ lead of contact $i$ at energy E, we expand the scattering state $|\psi^l\rangle$ of electrons that enters Eq.\ref{lippmann} in terms of atomic orbitals
\begin{multline}
|\psi^l\rangle =\sum_{n_l=-\infty}^{-1}(e^{ik^{l}n_la}+r_l e^{-ik^{l}n_la})|n_l\rangle\\+\sum_i C_i |c_i\rangle+\sum_{\beta} \sum_{n_\beta=1}^{\infty}t_{\beta,l}e^{ik^{\beta}n_\beta a}|n_\beta\rangle
\label{scatstate}
\end{multline}
where, $l$ specifies the 1D lead from which the electron is injected into the BLG nanostructure (belongs to one of the contacts in Fig.\ref{nanostructure}), $r_l $ is the amplitude for reflection of the electron back into the same 1D lead $l$, $i$ specifies the sites of the carbon atoms in the nanostructure, and $\beta$ specifies the 1D leads into which the electron is transmitted with amplitude $t_{\beta,l}$. By inserting the Eqs.\ref{scatstate}, \ref{green2}, \ref{Green} into the Eq.\ref{lippmann}, and applying $\langle c_i|, \langle {-1}_l|$ and $ \langle{+1}_\beta|$ to the result,  the Lippmann-Schwinger equation becomes a system of linear equations. This is solved numerically to find the quantum transmission amplitudes $t_{\beta,l}$ and reflection amplitudes $r_l $ that are required to obtain the transmission and reflection coefficients $T_{ji}$ and $R_{ii}$ that enter the B\"{u}ttiker Eq.\ref{buttiker} that is used to calculate the non-local resistance $R_{NL}$. The solution of the above system of linear equations  also yields
the coefficients $C_i$ in Eq. \ref{scatstate} that define the electronic transport states $|\psi^l\rangle$ within the BLG nanostructure.
\section{Valley-projected states}
In order to evaluate the valley currents in the bilayer graphene nanostructure, the scattering states of electrons $|\psi^l\rangle$ are projected onto crystal Bloch states, which can be written as
\begin{multline}
|\psi_{\bf k}^\alpha \rangle=\frac{1}{\sqrt{N}}\sum_{i=1}^Ne^{i\bf k . \bf R_i} (c^\alpha_{A_1}({\bf k} ) |p_{z,i}^{A_1}\rangle +c^\alpha_{B_1}({\bf k} ) |p_{z,i}^{B_1}\rangle+\\
c^\alpha_{A_2}({\bf k} ) |p_{z,i}^{A_2}\rangle+c^\alpha_{B_2}({\bf k} ) |p_{z,i}^{B_2}\rangle)
\end{multline}
where $\alpha = 1, ... 4$ denotes the different Bloch states having wave vector $\bf k$, $N$ is the total number of unit cells in the BLG nanostructure, ${\bf R}_i$ are the Bravais lattice vectors of bilayer graphene, and ${ |p_{z,i}^{A_1}\rangle, |p_{z,i}^{A_2}\rangle, |p_{z,i}^{B_1}\rangle, |p_{z,i}^{B_2}\rangle}$ are the carbon atomic orbitals in $i^{th}$ unit cell.  Then, the rhombic Brillouin zone of bilayer graphene represented in Fig.\ref{nanostructure} is divided into two parts and the Bloch state is assumed to belong to the valley K (K$'$) if its wave vector lies within the the upper (lower) half of the rhombic Brillouin zone.  Hence, the projected states are calculated as:
\begin{equation}
\begin{aligned}
&|\psi_K^l\rangle=A\sum_{\alpha,{\bf k}\in K}|\psi^\alpha_{\bf k}\rangle\langle\psi^\alpha_{\bf k}|\psi^l\rangle\\
&|\psi_{K'}^l\rangle=A\sum_{\alpha,{\bf k}\in K'}|\psi^\alpha_{\bf k}\rangle\langle\psi^\alpha_{\bf k}|\psi^l\rangle.
\end{aligned}
\label{proj}
\end{equation}
The Bloch states $|\psi_{\bf k}^\alpha\rangle$ are defined on a continuum in ${\bf k}$-space. However, in practice the sums in Eq.\ref{proj} are evaluated on a mesh of ${\bf k}$-points, the number of  ${\bf k}$-points in the mesh being chosen large enough for convergence of the valley currents being calculated with the help of the valley-projected states.  Thus appropriate normalization of the projected states is necessary. In Eq.\ref{proj}, $A=\frac{\text{number of unit cells}}{\text{number of mesh points}}$ is the required normalization factor, and the sums run over the mesh points belonging to the corresponding valley $K$ or $K'$. Note that the Bloch states of a bilayer graphene crystal with wave vector $\bf k$ can be written as linear combinations of Bloch states (with the same wave vector $\bf k$) of the two monolayer graphene crystals that comprise the bilayer. Therefore, for the purpose of projecting the scattering states of electrons $|\psi^l\rangle$ onto the subspace of crystal Bloch states with wave vector $\bf k$, Bloch states of a pair of decoupled graphene monolayer crystals can be used instead of Bloch states of the bilayer in Eq. B.1 and B.2, with equivalent results.
This projection method was used in the present work.
\end{appendices}

{

\end{document}